\documentclass[aps,prl,showpacs,twocolumn]{revtex4-1}
\usepackage{graphicx}
\usepackage{amsfonts,bbm,amssymb}
\usepackage{txfonts}
\usepackage{color}

\def\vmbb#1{\varmathbb{#1}}
\newcommand{\be}{\begin{equation}}
\newcommand{\ee}{\end{equation}}

\newcommand{\un}[1]{{{#1}'}}
\newcommand{\bra}[1]{{\langle #1 \vert}}

\newcommand{\ket}[1]{{\vert #1 \rangle}}

\newcommand{\ii}{ {\rm i} }
\newcommand{\dd}{ {\rm d} }

\newcommand{\ZZ}{\mathbb{Z}}

\newcommand{\CC}{\mathbb{C}}

\newcommand{\z}{{\rm z}}
\newcommand{\LL}{{\hat {\cal L}}}

\newcommand{\mm}[1]{{\mathbf{#1}}}

\def\tr{{\,{\rm tr}}}

\def\one{\mathbbm{1}}

\def\tr{\,{\rm tr}\,}
\def\ket#1{|#1\rangle}
\def\bra#1{\langle#1|}

\def\ii{{\rm i}}

\def\z{{\rm z}}

\def\etal#1{#1}

\def\tit#1{}

\newcommand{\half}{{\textstyle\frac{1}{2}}}

\begin{document}

\title{Infinitely dimensional Lax structure for the one-dimensional Hubbard model}

\author{Vladislav Popkov$^1,^2$ and Toma\v z Prosen$^3$}
\affiliation{$^1$ Institut f\" ur Theoretische Physik, Universit\" at zu K\" oln, D-50937 Cologne, Germany}
\affiliation{$^2$ CSDC Universit\`a di Firenze, via G. Sansone 1, 50019 Sesto Fiorentino, Italy}
\affiliation{$^3$ Faculty of Mathematics and Physics, University of Ljubljana, SI-1000 Ljubljana, Slovenia}

\date{\today}

\begin{abstract}
We report a two-parametric irreducible infinitely dimensional representation of the Lax integrability condition for the fermi Hubbard chain.
Besides being of fundamental interest, hinting on possible novel quantum symmetry of the model, our construction allows for an explicit representation of an exact steady state many-body density operator for non-equilibrium boundary-driven Hubbard chain with arbitrary (asymmetric) particle source/sink rates at the letf/right end of the chain and with arbitrary boundary values of chemical potentials.
\end{abstract}


\maketitle

{\em Introduction.--}
The  one-dimensional (1D) fermi Hubbard model is one of the key paradigms of exactly solvable models in theoretical physics \cite{book}.
Exact integrability is a remarkable exceptional property of a set of classical and quantum models,
playing a prominent role in theoretical physics. Integrability is essentially the only assumption-free handle on non-perturbative physics of strong interactions.
Exact solution of 2D Ising model formed a basis for the theory of phase transitions, solution of the Korteweg de Vries equation gave rise to soliton theory, solution of the $XYZ$ spin chain model showed a breakup of the scaling relations, and solution of 1D Hubbard model laid a milestone in understanding the problem of itinerant electrons in strongly correlated systems.
The integrability comprises an existence of a Yang Baxter structure, or Lax structure, consisting of R-matrix and Lax operators, depending on a free {\em spectral parameter}.
The Lax operator generally provides a crucial ingredient of integrability,
 an infinite set of mutually commuting local integrals of motion, through e.g. a
 construction of the commuting parameter-dependent transfer-matrices  \cite{korepin,sklyanin,grabowski}.
 The  one-dimensional Hubbard model stays aside in the list of integrable systems,
as, since its discovery, it continues to raise many deep fundamental questions.
Two decades after the discovery of the nested Bethe Ansatz for Hubbard model \cite{lieb,lieb2}
a Yang-Baxter formulation of the problem has been found  by Shastry \cite{shastry:86,shastry:88}.
Remarkably, Shastry's R-matrix has several unique properties \cite{book} not shared among most of other known integrable systems.

A recent progress has allowed to push the concepts of integrability of quantum systems beyond the thermal equilibrium:
exact nonequilibrium steady states (NESS) for
a set of quantum chains driven far from equilibrium by local incoherent markovian noise, were
calculated \cite{prosen:2011a, prosen:2011b, kps, kps2, pip13, iz, pi14,prosen:14a}. The non-equilibrium integrability, it turned out,
results from an additional degree of freedom, hidden in the standard Yang-Baxter structure for its `equilibrium' counterpart.
This additional degree of freedom is a {\em representation parameter} (being a complex number in non-equilibrium setting and connected
with the dissipative coupling at the model's boundaries), gives rise to
non-unitary representations of quantum symmetries of these models and allows to formulate a Matrix Product Ansatz
for a NESS. Moreover, these representations are infinite-dimensional, unlike their `equilibrium' counterparts.

Existence of an extra representation parameter in the Yang-Baxter  construction (alongside with the inherited
spectral parameter) not only allowed to find exact non-equilibrium generalizations of well-known integrable models,
but gave rise to an intrinsically new infinite set of  conservation laws for their equilibrium counterparts,
through a two-parameter commuting family  of operators, generated by  monodromy matrix elements.
Application of these new conserved quantities has allowed to solve a long-standing problem of high-temperature ballistic transport in the gapless anisotropic Heisenberg spin-1/2 ($XXZ$) chain
by establishing a rigorous lower bound on the corresponding spin Drude weight \cite{prosen:2011a,pi13,prosen:14c,affleck}. However, for the Hubbard model such a Lax representation was lacking.

Here we report a Lax structure for the Hubbard model, containing both, a free spectral parameter and an extra
(complex) representation parameter. As in previously solved examples ($XXZ$ model), it is infinite-dimensional
and gives rise to a two-parameter commuting family of operators generated by a monodromy matrix,
yielding a new set of conservation laws. For a fixed  value of the spectral parameter, the new
Lax structure yields an appealing factorization of our previous result \cite{prosen:14a}.
Remarkably, \textit{unlike} in the previously solved $XXZ$ model, the presented non-equilibrium Lax operator is different from
 the known one \cite{shastry:86} and cannot be reduced to the latter by means of auxiliary space truncation and gauge
transformations. Thus it is not a generalization (like in other integrable example \cite{pi13} where the non-equilibrium Lax operator can be encoded
within the so-called universal $R$-matrix of the quantum group $U_q(\mathfrak{sl}(2))$ symmetry of the $XXZ$ model), and it reveals a fundamentally different symmetry of the 1D Hubbard model. The new Lax solution  yields a spectrum of potential applications, from quantum transport in condensed matter to AdS/CFT duality and integrable ${\cal N}=4$ super Yang-Mills theory \cite{beisertreview}.
In the present communication we show how our generalised Lax operator can be employed to solve asymmetrically boundary driven Hubbard chains with arbitrary source/sink rates and boundary chemical potentials, where the non-vanishing value of the spectral parameter corresponds to the asymmetry of the driving.


We consider a fermi Hubbard chain on $n$ sites formulated in terms of a spin $1/2$ ladder.
Let ${\cal H}_{\rm p} = \CC^2 \otimes \CC^2$ be a {\em local physical Hilbert space} supporting two independent sets of Pauli matrices
$\sigma^s,\tau^t$, $s,t\in{\cal J}:=\{+,-,0,\z\}$, $\sigma^0=\tau^0=\one$. Embedding ${\cal H}_{\rm p}$ into the Hilbert space of an $n$-site ladder ${\cal H}_{\rm p}^{\otimes}$, one defines
local spin operators $\sigma^s_j,\tau^t_j$, for $j\in\{1,\ldots,n\}$.
The Hubbard hamiltonian with arbitrary boundary chemical potentials $\mu_{\rm L/R}$ \cite{notechem} then reads
\begin{eqnarray}
H &=& \sum_{j=1}^{n-1} h_{j,j+1} + h_{\rm L} + h_{\rm R}, \label{eq:H}\\
h_{1,2} &:=& h^\sigma_{1,2} + h^\tau_{1,2} + \frac{u}{2}\left(\sigma^\z_1 \tau^\z_1 + \sigma^\z_{2} \tau^\z_{2}\right), \label{eq:h12} \\
h_{\rm L/R} &:=& \frac{u}{2} \sigma^\z_{1/n} \tau^\z_{1/n} + \frac{\mu_{\rm L/R}}{2}\left(\sigma^\z_{1/n} + \tau^\z_{1/n}\right), \label{eq:hLR}
\end{eqnarray}
where $h^\sigma_{1,2}:=2 \sigma^{+}_1 \sigma^-_2 + 2 \sigma^{-}_1 \sigma^+_2$,
$h^\tau_{1,2}:=2 \tau^{+}_1 \tau^-_2 + 2 \tau^{-}_1 \tau^+_2$ are the free hoping operators of the respective particle species. Note that $(1,2)$ designates a generic pair of neighboring sites.
Dimensionless interaction parameter
$u=U/(2t_{\rm h})$ contains standard Hubbard interaction $U$ and hopping amplitude $t_{\rm h}$.
The standard fermionic Hamiltonian $H=-\sum_{j,s} (c^\dagger_{s,j}c_{s,j+1}+{\rm H.c.}) +  2u \sum_{j} (n_{\uparrow,j}-\frac{1}{2})(n_{\downarrow,j}-\frac{1}{2})+\mu_{\rm L}(n_{\uparrow,1}+n_{\downarrow,1}-1)+\mu_{\rm R}(n_{\uparrow,n}+n_{\downarrow,n}-1)$ is reconstructed via Jordan-Wigner transformation $c_{\uparrow,j}=P^{(\sigma)}_{j-1} \sigma_j^-$ and $c_{\downarrow,j}=P^{(\sigma)}_n P^{(\tau)}_{j-1} \tau_j^-$, $n_{s,j} := c^\dagger_{s,j}c_{s,j}$,
 where $P^{(\sigma)}_j:=\sigma_1^{\rm z} \cdots \sigma_j^{\rm z}$, $P^{(\tau)}_j:=\tau_1^{\rm z} \cdots \tau_j^{\rm z}$. Let us define the spin-flip (or particle-hole) operator $G$, i.e. permutation between
$\sigma$ and $\tau$ spins (or fermion species), as $G \sigma^s G = \tau^s$, $G^2=\one$. Clearly, $G h^{\sigma}_{1,2}G = h^{\tau}_{1,2}$, and $G h_{1,2} G = h_{1,2}$, $G H G = H$.

\begin{figure}
\begin{center}
\includegraphics[width=\columnwidth]{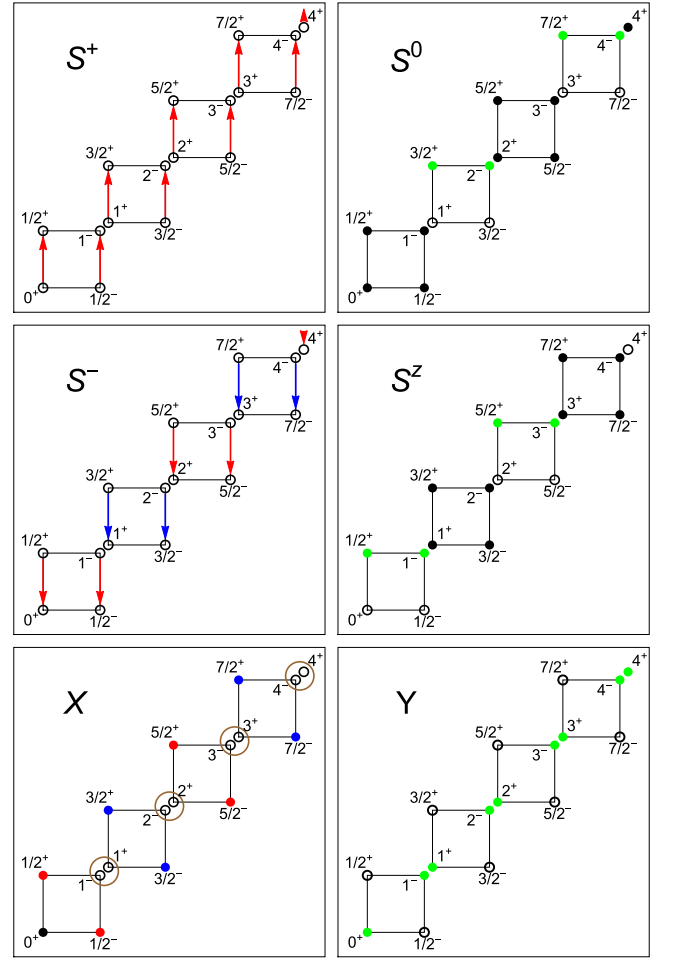}
\vspace{-8mm}
\end{center}
\caption{(Color online) Diagrammatic representation of factors of the Lax operator where auxiliary states are labelled by vertices ${\cal V}$. Diagrams for $\mm{T}^s$ are obtained by
reflection of those of $\mm{S}^s$ across the diagonal.
Red/blue arrows indicate offdiagonal transitions with amplitude $\pm \sqrt{2}$.
Red, blue, green, black, open points represent diagonal multiplications by $\omega,-\omega,\propto\lambda,1,0$, respectively, and brown circles represent multiplications by $2\times 2$ matrices $X_k$.}
\label{ST}
\end{figure}

{\em The Lax operator.--}
We introduce an infinitely dimensional {\em auxiliary Hilbert space},
${\cal H}_{\rm a} = {\rm LinearSpan}\{\ket{p}, p\in{\cal V}\}$,
whose orthonormal basis is conveniently labelled by vertices of a graph (see Fig.~\ref{ST}),
${\cal V}=\{0^{+},\frac{1}{2}^{+},\frac{1}{2}^{-},1^{-},1^{+},\frac{3}{2}^{+},\frac{3}{2}^{-},2^{-},2^{+}\ldots\}$.
We extend the definition of the spin-flip $\mm{G}$ over ${\cal H}_{\rm a}$ as a diagonal reflection of the graph, $\mm{G}\ket{k^\pm} = \ket{k^\pm}$, $\mm{G} \ket{k\!+\!\half^\pm} = \ket{k\!+\!\half^\mp}$, $k \in \ZZ^+$.
Here and below we shall use bold-roman letters to designate operators which are non-scalar over ${\cal H}_{\rm a}$. We begin our analysis with a simple observation:

\smallskip
\noindent
{\em Lemma:} Assume there exist operators $\mm{S},\acute{\mm{S}},\grave{\mm{S}},\mm{T},\acute{\mm{T}},\grave{\mm{T}} \in {\rm End}({\cal H}_{\rm a}\otimes {\cal H}_{\rm p})$, and $\mm{X},\mm{Y}\in {\rm End}({\cal H}_{\rm a})$ (acting as scalars over ${\cal H}_{\rm p}$),  satisfying
\begin{eqnarray}
&& [h^\sigma_{1,2},\mm{S}_1 \mm{X} \mm{S}_2]  = \acute{\mm{S}}_1 \mm{X} \mm{S}_2 - \mm{S}_1\mm{X} \grave{\mm{S}}_2, \label{eq:id1}\\
&& [h^\tau_{1,2},\mm{T}_1 \mm{X} \mm{T}_2] = \acute{\mm{T}}_1 \mm{X} \mm{T}_2 - \mm{T}_1\mm{X}\grave{\mm{T}}_2, \label{eq:id2}\\
&& \mm{S}\acute{\mm{T}} + \mm{T}\acute{\mm{S}} - \grave{\mm{S}}\mm{T} - \grave{\mm{T}}\mm{S} = [\mm{Y}-u \sigma^\z \tau^\z,\mm{S}\mm{T}], \label{eq:id3}\\
&& [\mm{S},\mm{T}] = 0, \label{eq:id4}\\
&&[\mm{X},\mm{Y}] = 0. \label{eq:id5}
\end{eqnarray}
Subscripts, like in $\mm{S}_j$, indicate independent local physical spaces pertaining to sites $j$. Then, one can define a Lax operator and its `derivative'
$\mm{L},\tilde{\mm{L}}\in {\rm End}({\cal H}_{\rm a}\otimes {\cal H}_{\rm p})$ as
\begin{eqnarray}
&&\mm{L} = \mm{S}\mm{T}\mm{X}, \label{eq:L} \\
&&\tilde{\mm{L}} = \half(\mm{S}\acute{\mm{T}} + \mm{T}\acute{\mm{S}} + \grave{\mm{S}}\mm{T} + \grave{\mm{T}}\mm{S} - \{\mm{Y},\mm{S}\mm{T}\})\mm{X}, \label{eq:Lt}
\end{eqnarray}
such that a so-called Sutherland-Shastry (or generalized \cite{shastry:86,massarani} local operator divergence \cite{sutherland} gLOD) condition holds
\begin{equation}
[h_{1,2},\mm{L}_1 \mm{L}_2] = (\tilde{\mm{L}}_1 + \mm{Y} \mm{L}_1)\mm{L}_2 - \mm{L}_1(\tilde{\mm{L}}_2 + \mm{L}_2\mm{Y}).
\label{eq:gLOD}
\end{equation}
The proof is a straightforward insertion of (\ref{eq:L},\ref{eq:Lt}) into Eq.~(\ref{eq:gLOD}) followed by subsequent application of identities (\ref{eq:id1}-\ref{eq:id5}) observing the definition (\ref{eq:h12}).

 We continue by {\em deriving} an explicit closed form representation of algebraic identities (\ref{eq:id1}-\ref{eq:id5}). Assuming the spin-flip symmetry
 \begin{equation}
 \mm{G} \mm{S} \mm{G} = \mm{T},\,
 \mm{G} \acute{\mm{S}} \mm{G} = \acute{\mm{T}},\,
 \mm{G} \grave{\mm{S}} \mm{G} = \grave{\mm{T}},\, [\mm{G},\mm{X}]=[\mm{G},\mm{Y}]=0,
 \label{eq:sym}
 \end{equation}
 and writing out the components $\mm{S}=\sum_{s\in{\cal J}} \mm{S}^s \sigma^s$, $\mm{T}=\sum_{t\in{\cal J}} \mm{T}^t \tau^t$, and similarly for $\acute{\mm{S}},\grave{\mm{S}},\acute{\mm{T}},\grave{\mm{T}}$, we find that Eqs. (\ref{eq:id1}) and (\ref{eq:id2}) are equivalent, Eq.~(\ref{eq:id3}) is invariant under $\mm{G}$, while Eq.~(\ref{eq:id4}) implies $[\mm{S}^s,\mm{T}^t]\equiv 0$.

Eqs. (\ref{eq:id1},\ref{eq:id2}) are in fact just a particularly `decorated' 6-vertex Yang-Baxter equations for free fermion (or $XX$) chains. We shall thus make an ansatz for $\mm{S}^s,\mm{T}^t$
in which each square plaquette $\{ k^+, k\!+\!\half^+,k\!+\!\half^-,k\!+\!1^-\}$ of the graph spans a pair of representations of a free fermion algebra (see Fig.~\ref{ST}), namely requiring that
$\{\mm{S}^+,\mm{S}^-\}$ (and similarly for $\{\mm{T}^+,\mm{T}^-\}$ via (\ref{eq:sym})) is  in the center of the algebra generated by $\mm{S}^s,\mm{T}^t$
\begin{equation}
[\{\mm{S}^+,\mm{S}^-\},\mm{S}^s] = [\{\mm{S}^+,\mm{S}^-\},\mm{T}^t] = 0,\quad s,t\in{\cal J}.
\end{equation}
One finds that these conditions are fulfilled by an ansatz
\begin{eqnarray}
&&\mm{S}^+ =\sqrt{2} \sum_{k=0}^\infty \left(\ket{k^+}\bra{k\!+\!\half^+} + \ket{k\!+\!\half^+}\bra{k\!+\!1^-}\right),\label{eq:Sp} \\
&&\mm{S}^-  = \sqrt{2} \sum_{k=0}^\infty (-1)^k \left(\ket{k\!+\!\half^+}\bra{k^+} + \ket{k\!+\!1^-}\bra{k\!+\!\half^+}\right),\nonumber\\
&&\mm{S}^0 = \sum_{k=0}^\infty \bigl(\ket{2k^+}\bra{2k^+} + \ket{2k\!+\!\half^+}\bra{2k\!+\!\half^+} \nonumber \\
&&\qquad+ \ket{2k\!+\!1^-}\bra{2k\!+\!1^-}+ \ket{2k\!+\!\half^-}\bra{2k\!+\!\half^-}\bigr)\nonumber\\
&&\quad+ \lambda\sum_{k=1}^\infty \left( \ket{2k\!-\!\half^+}\bra{2k\!-\!\half^+} + \ket{2k^-}\bra{2k^-}\right), \nonumber \\
&&\mm{S}^\z = \sum_{k=1}^\infty \bigl(\ket{2k\!-\!1^+}\bra{2k\!-\!1^+} + \ket{2k\!-\!\half^+}\bra{2k\!-\!\half^+} \nonumber \\
&&\qquad + \ket{2k^-}\bra{2k^-} + \ket{2k\!+\!\half^-}\bra{2k\!+\!\half^-}\bigr)  \nonumber\\
&&\quad + \lambda\sum_{k=0}^\infty\left(\ket{2k\!+\!\half^+}\bra{2k\!+\!\half^+} + \ket{2k\!+\!1^-}\bra{2k\!+\!1^-}\right), \nonumber
\end{eqnarray}
where $\lambda\in\CC$ is a free parameter.
Eqs. (\ref{eq:sym}) imply definition of another set of auxilliary fermi operators $\mm{T}^t=\mm{G}\mm{S}^t\mm{G}$, such that Eq.~(\ref{eq:id4}) is satisfied.

Further, one finds a remarkably consistent ansatz for the `interaction' operator $\mm{X}$ coupling the neighboring plaquettes:
\begin{eqnarray}
\mm{X} &=& \ket{0^+}\bra{0^+} +
 \sum_{k=1}^\infty (-1)^{k}\!\!\!\sum_{\nu,\nu'\in\{-,+\}}   \ket{k^{\nu}} X^{\nu,\nu'}_k  \bra{k^{\nu'}} \nonumber\\
&+& \omega \sum_{k=0}^\infty (-1)^k \left(\ket{k\!+\!\half^+}\bra{k\!+\!\half^+} +
\ket{k\!+\!\half^-}\bra{k\!+\!\half^-}\right),  \label{eq:Xansatz}
\end{eqnarray}
where $X_k=\{ X^{\nu,\nu'}_k \}_{\nu,\nu'\in\{-,+\}}$ are still unknown $2\times 2$ matrices and $\omega\in\CC$ is another free parameter.
Namely, Eq.~(\ref{eq:id1}) yields a system of linear equations for auxiliary operators $\acute{\mm{S}}^s\mm{X},\mm{X}\grave{\mm{S}}^s$,
with a unique solution parametrised by $X_k,\omega,\lambda$:
\begin{eqnarray}
&&\acute{\mm{S}}^+\mm{X}  =  -2\sqrt{2}\sum_{k=1}^{\infty}(-1)^k X^{+-}_k \ket{k^-}\bra{k\!+\!\half^+}, \label{eq:Sb}\\
&&\acute{\mm{S}}^-\mm{X} =   -2\sqrt{2}\sum_{k=1}^{\infty} X^{-+}_k\ket{k^+}\bra{k\!-\!\half}, \nonumber\\
&&\mm{X}\grave{\mm{S}}^{+}  =  2\sqrt{2}\sum_{k=1}^{\infty}
(-1)^k X^{+-}_k\ket{k\!-\!\half^-}\bra{k^+} \nonumber\\
&&\mm{X}\grave{\mm{S}}^- =  -2\sqrt{2}\sum_{k=1}^{\infty}
X^{-+}_k \ket{k\!+\!\half^+}\bra{k^-}, \nonumber\\
&&\acute{\mm{S}}^0\mm{X}=\mm{X}\grave{\mm{S}}^0=
2\sum_{k=1}^{\infty}\bigl(\omega\ket{2k\!-\!1^+}\bra{2k\!-\!1^+}-\omega\ket{2k^-}\bra{2k^-} \nonumber \\
&&\quad - X^{++}_{2k-1}\ket{2k\!-\!\half^+}\bra{2k\!-\!\half^+}-X^{--}_{2k}\ket{2k\!-\!\half^-}\bra{2k\!-\!\half^-}\bigr) \nonumber \\
&&+2\lambda
\sum_{k=0}^{\infty}
\bigl(-\omega \ket{2k^+}\bra{2k^+}+X^{--}_{2k+1}\ket{2k\!+\!\half^-}\bra{2k\!+\!\half^-}\bigr), \nonumber\\
&&\acute{\mm{S}}^\z\mm{X}=\mm{X}\grave{\mm{S}}^\z=
 2\sum_{k=0}^{\infty}\bigl(\omega \ket{2k\!+\!1^-}\bra{2k\!+\!1^-}-\omega\ket{2k^+}\bra{2k^+} \nonumber \\
 &&\quad + X^{++}_{2k}\ket{2k\!+\!\half^+}\bra{2k\!+\!\half^+} + X^{--}_{2k+1}\ket{2k\!+\!\half^-}\bra{2k\!+\!\half^-}\bigr) \nonumber \\
 && + 2\lambda
\sum_{k=1}^{\infty}\bigl(\omega\ket{2k\!-\! 1^+}\bra{2k\!-\! 1^+}-X^{--}_{2k} \ket{2k\!-\!\half^-}\bra{2k\!-\!\half^-}\bigr). \nonumber
\end{eqnarray}
Assuming $\mm{X}$ to be invertible (i.e., $\omega\neq 0$, $\det X_k \neq 0$) and
plugging expressions (\ref{eq:Sb}) to the remaining identity (\ref{eq:id3}) result in (i) a unique consistent expression for the `spectral' operator $\mm{Y}$
\begin{equation}
\mm{Y}=-2\lambda u\sum_{k=0}^{\infty}\bigl(\ket{k^+}\bra{k^+} + \ket{k\!+\!1^-}\bra{k\!+\!1^-}\bigr),
\end{equation}
which clearly commutes with $\mm{X}$, as required by (\ref{eq:id5}),
and (ii) recurrence relations for the matrix elements of $X_k$:
$X^{--}_{k+1}=X^{--}_{k} - u \omega$, $X^{++}_{k+1}=X^{++}_{k} - u \omega (1-\lambda^2)$,
and $\det X_k = -\omega^2$, while also fixing the initial condition $X^{++}_{0}=1$, $X^{--}_{0}=-\omega^2$, yielding
\begin{equation}
X_k(\lambda,\omega)=\pmatrix{
-(\omega+k u)\omega & 1-(\omega+k u)\omega(1-\lambda^2)\cr
 -k u \omega & 1- k u \omega (1-\lambda^2)
}.
\label{Xk}
\end{equation}
Note that $X^{-+}_k/X^{+-}_k$ can be chosen freely exploring a gauge freedom
$\ket{k^\pm}\to \xi^{\pm 1} \ket{k^\pm}$, $k=1,2\ldots$
We have thus constructed two-parameter representation of the Lax operator $\mm{L}(\lambda,\omega)=\mm{S}(\lambda)\mm{T}(\lambda)\mm{X}(\lambda,\omega)$ satisfying gLOD (\ref{eq:gLOD}). We propose to call $\lambda$ a {\em spectral parameter} and $\omega$ a {\em representation parameter}. Remarkably, our representation is generically of infinite-dimension, for any nonzero $u$, and can only be truncated to $4k$-dim.~span of first $k$ plaquette states at special points along algebraic curves $X^{-+}_k(\lambda,\omega) = 0$ in the $\lambda-\omega$ plane.
On the other hand, we have checked that the Shastry's Lax operator \cite{shastry:86}, together with appropriate auxiliary operators, form a $4$-dim. representation of the algebra (\ref{eq:id1}-\ref{eq:id5}) as well.
So one can be tempted to think that our Lax structure,  for $X^{-+}_1(\lambda,\omega(\lambda))=0$,  and Shastry's Lax 
structure can be equivalent; however, so far, we could not find any correspondence.

{\it Lax form of NESS for asymmetric boundary driving.--}
As an application of the novel Lax operator we consider a markovian master equation
$\dd\rho_t/\dd t = \hat{\cal L}\rho_t$ for an open Hubbard chain with Hamiltonian (\ref{eq:H}) and driven by pure source/sink at the left/right ends with the non-negative rates $\Gamma_{\rm L/R}$:
\begin{equation}
\hat{\cal L}\rho=-\ii [H,\rho]+
\bigl(\Gamma_{\rm L}(\hat{\cal D}_{\sigma^+_1}+\hat {\cal D}_{\tau^+_1})
+
\Gamma_{\rm R}(\hat{\cal D}_{\sigma^-_n}+\hat {\cal D}_{\tau^-_n})\bigr)\rho,
\end{equation}
where $\hat{\cal D}_L \rho = 2 L \rho L^\dagger - \{L^\dagger L,\rho\}$ is a Lindblad dissipator \cite{lin,gks}, a linear map over ${\rm End}({\cal H}^{\otimes n}_{\rm p})$.
We shall now prove the following explicit result on NESS $\rho_{\infty}$:

\smallskip
\noindent
{\em Theorem:}
The unique \cite{note} fixed point $\LL\rho_\infty=0$ of boundary driven Hubbard chain reads
\begin{equation}
\rho_\infty = (\!\tr R)^{-1} R,\quad R = \Omega\,\Omega^\dagger M,
\end{equation}
where $\Omega=\Omega(\lambda,\omega)$ is a {\em highest-weight} transfer matrix
\begin{equation}
\Omega = \bra{0^+} \mm{L}_1(\lambda,\omega)\mm{L}_2(\lambda,\omega)\cdots  \mm{L}_n(\lambda,\omega)\ket{0^+}
\end{equation}
and $M$ is a diagonal operator
\begin{equation}
M = M_1 M_2\cdots M_n,\quad M_j = \exp\left(\eta (\sigma^\z_j+\tau^\z_j)\right)
\end{equation}
with $\eta = \half\log \Gamma_{\rm L}/\Gamma_{\rm R}$ and parameters $\lambda,\omega$ are related to coherent and incoherent biases
\begin{equation}
\lambda = \frac{\Gamma_{\rm L}-\Gamma_{\rm R} - \ii (\mu_{\rm L}+\mu_{\rm R})}{\Gamma_{\rm L}+\Gamma_{\rm R} -
\ii (\mu_{\rm L}-\mu_{\rm R})},\;
\omega =\frac{1}{4}\left(\mu_{\rm L}-\mu_{\rm R} + \ii\left(\Gamma_{\rm L}+\Gamma_{\rm R}\right)\right).
\label{eq:pars}
\end{equation}
{\em Proof.} Let us now invoke two copies of the auxiliary space and define operators $\mm{S},\mm{T},\mm{\un{S}},\mm{\un{T}}\in{\rm End}({\cal H}_{\rm a}\otimes {\cal H}_{\rm a}\otimes {\cal H}_{\rm p})$ as
\begin{eqnarray*}
&&\mm{S} = \sum_s \mm{S}^s \otimes \one_{\rm a} \otimes \sigma^s,\;\,\qquad \mm{T} = \sum_t \mm{T}^t \otimes \one_{\rm a} \otimes \tau^t,\quad {\rm and}\\
&&\mm{\un{S}} = \sum_s  \one_{\rm a} \otimes \mm{\bar{S}}^s \otimes (\sigma^s)^T,\quad \mm{\un{T}} = \sum_t \one_{\rm a} \otimes \mm{\bar{T}}^t \otimes (\tau^t)^T.
\end{eqnarray*}
$()^T$ denotes the matrix transposition and $\bar{\mm{S}}$ the complex conjugation, i.e. replacement $\lambda,\omega,\to\bar{\lambda},\bar{\omega}$, and similarly for
$\acute{\mm{S}},\grave{\mm{S}},\acute{\mm{\un{S}}},\grave{\mm{\un{S}}},\acute{\mm{T}},\grave{\mm{T}},\acute{\mm{\un{T}}},\grave{\mm{\un{T}}}$,
and $\mm{X},\mm{\un{X}},\mm{Y},\mm{\un{Y}}\in{\rm End}({\cal H}_{\rm a}\otimes{\cal H}_{\rm a})$.
In fact, the primed operators
$ \mm{\un{S}},\acute{\mm{\un{S}}},\grave{\mm{\un{S}}},\mm{\un{T}},\acute{\mm{\un{T}}},\grave{\mm{\un{T}}},\mm{\un{X}},\mm{\un{Y}} $ generate a {\em conjugate} representation of the algebra (\ref{eq:id1}-\ref{eq:id5}). Noting $[h_{1,2},M_1 M_2]=0$ and the Jacobi identity one finds that the following double auxiliary operators
\begin{equation}
\vmbb{L}_j = \mm{L}_j\mm{\un{L}}_{\!\!j} M_j,\;
\tilde{\vmbb{L}}_j = (\tilde{\mm{L}}_j \mm{\un{L}}_{\!\!j} - \mm{L}_j\tilde{\mm{\un{L}}}_{\!\!j})M_j,\;
\vmbb{Y} = \mm{Y}-\mm{\un{Y}},
\end{equation}
also respect  gLOD (\ref{eq:gLOD}), resulting in the telescoping series
\begin{eqnarray}
\sum_{j=1}^{n-1} \left[h_{j,j+1}, \vmbb{L}_1\vmbb{L}_2\cdots \vmbb{L}_n\right] &=&
(\tilde{\vmbb{L}}_1+\{\vmbb{Y},\vmbb{L}_1\})\vmbb{L}_2\cdots\vmbb{L}_{n} \nonumber \\
&-&\vmbb{L}_1\cdots\vmbb{L}_{n-1}(\tilde{\vmbb{L}}_n+\{\vmbb{Y},\vmbb{L}_n\}).\qquad \label{eq:telescop}
\end{eqnarray}
Double Lax operator expresses NESS in a compact form
\begin{equation}
R = \bra{0^+,0^+} \vmbb{L}_1\vmbb{L}_2 \cdots \vmbb{L}_n \ket{0^+,0^+},
\end{equation}
hence the fixed point condition $\LL R = 0$ becomes, after applying (\ref{eq:telescop}) to $[H,R]$, equivalent
to a pair of equations for ultralocal operators at the boundary physical sites
\begin{eqnarray}
&& \bra{0^+,0^+}\left(\ii \Gamma_{\rm L}(\hat{\cal D}_{\sigma^+}+ \hat{\cal D}_{\tau^+})\vmbb{L} + \tilde{\vmbb{L}}+ \vmbb{L}\vmbb{Y}+[h_{\rm L},\vmbb{L}]\right)=0,\quad
\nonumber\\
&&\left(\ii \Gamma_{\rm R}(\hat{\cal D}_{\sigma^-}+ \hat{\cal D}_{\tau^-})\vmbb{L} - \tilde{\vmbb{L}} - \vmbb{Y}\vmbb{L}+[h_{\rm R},\vmbb{L}]\right)\ket{0^+,0^+}=0,\quad
\label{eq:bc}
 \end{eqnarray}
 where boundary interactions with fields, $h_{\rm L/R}$, are defined in (\ref{eq:hLR}).
 Using explicit forms (\ref{eq:Sp}-\ref{Xk}) and in particular $\mm{X}\ket{0^+,0^+}=\mm{\un{X}}\ket{0^+,0^+}=\ket{0^+,0^+}$, each of Eqs.~(\ref{eq:bc})
  results in $\dim{\cal H}_{\rm p}\times \dim{\cal H}_{\rm p}=16$ equations for (bra/ket) vectors from ${\cal H}_{\rm a}\otimes{\cal H}_{\rm a}$, most of them
  trivially satisfied, whereas the non-trivial ones being equivalent to conditions (\ref{eq:pars}).

 {\em Discussion.--} We have derived an infinitely dimensional irreducible representation of Lax operator and Sutherland-Shastry compatibility condition and shown how it can be employed to yield exact NESS of asymmetrically boundary driven Hubbard chain with arbitrary boundary chemical potentials.
One is thus able to explicitly compute physical observables in NESS. For example, linear dependence of the amplitudes (\ref{Xk}) on auxiliary state $k$ immediately yields, similarly as in Heisenberg chain \cite{prosen:2011b}, a universal scaling of the spin/charge currents $J\sim n^{-2}$ and cosine-shaped
spin/charge density profile, as observed in numerical simulations \cite{pz12} (details to be given elsewhere).
We are convinced that our fundamental result shall find applications far beyond the treatment of boundary driven quantum master equation. For example, computer algebra (without a proof yet!) suggests an existence of an infinitely dimensional intertwiner (R-matrix) between a pair of auxiliary spaces, implying exact commutativity \cite{pip13} of a two-parameter family of non-Hermitian transfer operators
\begin{equation}
[\Omega(\lambda,\omega),\Omega(\lambda',\omega')]=0,\quad\forall \lambda,\lambda',\omega,\omega'\in\CC.
\end{equation}
We note that our novel Lax operator provides an appealing factorisation of our previous result \cite{prosen:14a} -- derived by a more brut-force approach -- in the special case of zero spectral parameter $\lambda=0$.
We note that $\partial_\omega \Omega(\lambda,\omega)|_{\lambda=\omega=0}$ matches with the generator of Yangian symmetry of the Hubbard model \cite{uglov} truncated to a finite open chain of $n$ sites. However, the relevance and facility of other conservation laws derived from $\Omega(\lambda,\omega)$, say for establishing rigorous bounds on transport coefficients \cite{prosen:14b}, remain
an exciting problem to study in future. We stress that our attempts to link our novel concepts to the $4-$dim.~Lax matrix constructed by Shastry \cite{shastry:88}, or to quantum- or super-symmetries of the
related Hubbard-like models within the framework of integrable ${\cal N}=4$ super Yang-Mills theory \cite{koroteev, beisert}, failed so far. It is thus in our opinion an urgent question to establish whether these results imply existence of novel quantum symmetries of the 1D Hubbard model.

We thank E. Ilievski and G. M. Sch\" utz for stimulating discussions and acknowledge support by Deutsche Forschungsgemeinschaft (DFG)
and by the grants P1-0044, J1-5439, N1-0025 of Slovenian Research Agency (ARRS). We also thank
the Galileo Galilei Institute for Theoretical Physics, Florence, where part of this work was done,
for hospitality and for partial support.


\begin{thebibliography}{1}
\bibitem{notechem}
Note that a uniform bulk chemical potential term can be trivially removed from the discussion.
\end{thebibliography}
\end{document}